\documentclass{article}

\usepackage[preprint]{neurips_2025}

\usepackage{graphicx} 
\usepackage{natbib}          
\usepackage{amsmath,amsthm}
\usepackage{amsfonts} 
\usepackage{mathtools}
\usepackage{microtype} 
\usepackage{float} 
\usepackage{booktabs,threeparttable,tabularx}
\usepackage[hidelinks]{hyperref}
\usepackage[capitalise,noabbrev]{cleveref}
\usepackage{siunitx}
\usepackage{xfp}
\usepackage{catchfile}

\crefname{fact}{stylized fact}{stylized facts}
\Crefname{fact}{Stylized Fact}{Stylized Facts}
\crefname{assumption}{assumption}{assumptions}
\Crefname{assumption}{Assumption}{Assumptions}

\newtheorem{lemma}{Lemma}              
\newtheorem{prop}{Proposition}         
\newtheorem{fact}{Stylized Fact}       

\usepackage{authblk} 

\title{Forecasting AI Time Horizon Under Compute Slowdowns}

\author[1]{Parker Whitfill}
\author[2]{Ben Snodin}
\author[3]{Joel Becker}
\affil[1]{MIT} \affil[2]{Independent} \affil[3]{METR}

\date{November 2025}

\begin{document}

\maketitle

\begin{abstract}
METR's time horizon metric has grown exponentially since 2019, along with compute. However, it is unclear whether compute scaling will persist at current rates through 2030, raising the question of how possible compute slowdowns might impact AI agent capability forecasts. Given a model of time horizon as a function of training compute and algorithms, along with a model of how compute investment spills into algorithmic progress (which, notably, precludes the possibility of a software-only singularity), and the empirical fact that both time horizon and compute have grown at constant rates over 2019--2025, we derive that time horizon growth must be proportional to compute growth. We provide additional, albeit limited, experimental evidence consistent with this theory. We use our model to project time horizon growth under OpenAI's compute projection, finding substantial projected delays in some cases. For example, 1-month time horizons at $80\%$ reliability occur $7$ years later than simple trend extrapolation suggests. 
\end{abstract}

\providecommand{\keywords}[1]{\noindent\textbf{Keywords:} #1}

\section{Introduction}\label{sec:intro}
\citet{kwa2025measuring} show that, from 2019 to 2025, the time horizon of software tasks that AI agents can accomplish with 50\% accuracy has grown exponentially, with a steady 7-month doubling time. 
However, this trend is arguably driven by exponential growth in compute over the same period \citep{epoch2024trainingcomputeoffrontieraimodelsgrowsby45xperyear}. This research note provides a framework for forecasting METR's time horizon trend given potential compute slowdowns \citep{epoch2025computescalingwillslowdownduetoincreasingleadtimes, epoch2024canaiscalingcontinuethrough2030}, conditional on no software-only singularity.

We begin in \Cref{sec:theory} with a simple theoretical model in which time horizon depends on compute and algorithms, and algorithmic progress follows an ideas-production model where the input is experimental compute. (Note that this assumption is contestable, and precludes the possibility of a software-only singularity \citep{davidson2025willcomputebottleneckspreventasoftwareintelligenceexplosion}.)  We show that, given this model, the basic facts that time horizon and (experimental + training) compute have been growing at constant rates imply that the time horizon growth rate is causally proportional to the compute growth rate --- if the growth rate of compute halves, the growth rate of time horizon will also halve.

In \Cref{sec:empirics}, we test an empirical prediction of the theoretical model by comparing how log training compute relates to log time horizon within the Llama $3.1$ and Qwen $2.5$ families. We show that the relationship is approximately linear and the same across model families with different algorithms, consistent with the theory. This analysis, along with our theoretical framework, enables us to make an estimate of algorithmic progress in \Cref{sec:algorithmic-progress} --- approximately $3.5\times$ per year, consistent with \citet{ho2024algorithmic}.

\begin{figure}[H]
  \centering
  \includegraphics[width=\textwidth]{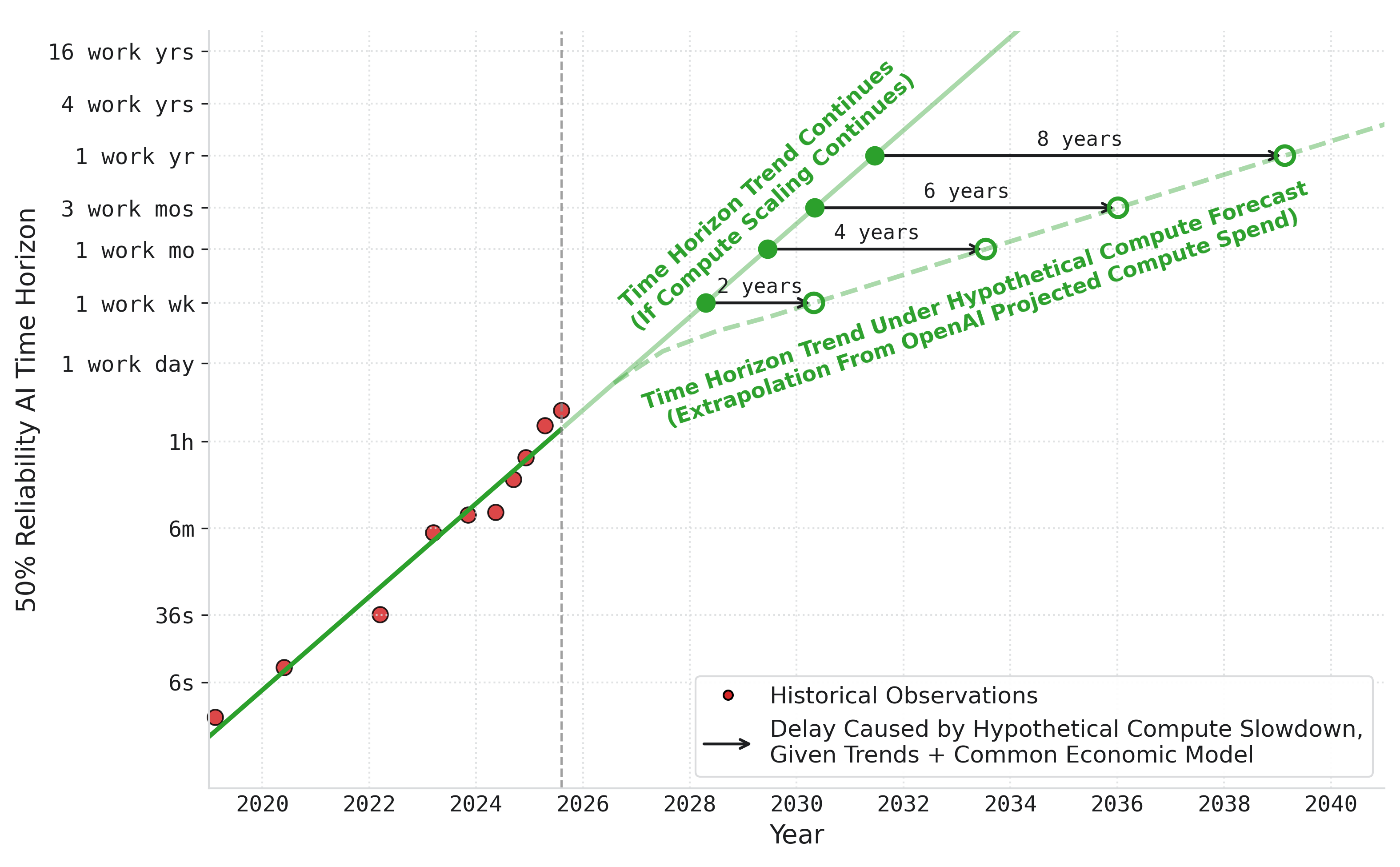}
  \caption{50\% reliability time horizon forecast implied by our model.}
  \label{fig:fig1}
\end{figure}

Lastly, we combine our model with a compute forecast to see how much delay a compute slowdown would cause. In particular, we construct a time series of OpenAI's compute spend from 2018--2025 and use OpenAI's own forecasts of their compute spend up to 2030 as reported in \cite{EfratiMuppidi2025OpenAI350B}. Figure \ref{fig:fig1} shows the effect of taking into account compute slowdowns for time horizon. While this scenario is primarily illustrative, it shows that some milestones can be substantially delayed --- for example, 1-month, $50$\% reliable AI arrives in 2033 instead of 2029. In general, we find that delays are more severe for higher time horizons and higher reliability thresholds.

\section{Theory}\label{sec:theory}
The goal of this section is to formally derive the relationship between compute growth and time horizon growth using the following two stylized facts. The first stylized fact comes directly from \citet{kwa2025measuring}.

\begin{fact}\label{fact:time}
    Time horizon is growing at a constant rate.
\end{fact}

The next stylized fact is that total R\&D compute spend is growing at a constant rate.  
\begin{fact}\label{fact:compute}
    Total R\&D compute is growing at a constant rate.
\end{fact}
The best available data for this stylized fact is for training compute \citep{epoch2024trainingcomputeoffrontieraimodelsgrowsby45xperyear}, but total R\&D compute (experimental plus training compute) also appears to be growing at an exponential rate. 

\begin{figure}[H]
  \centering
  \includegraphics[width=\linewidth]{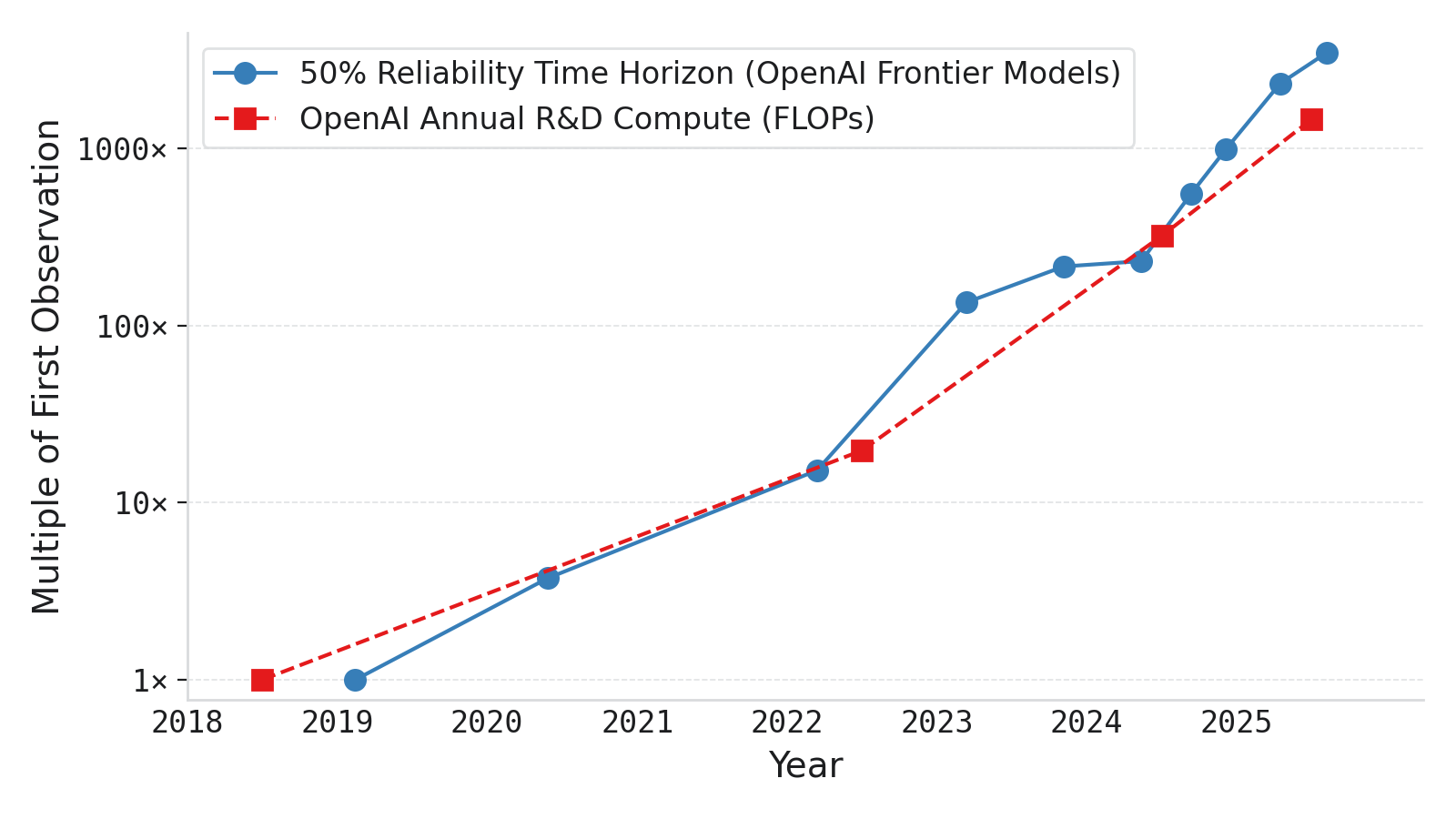}
  \caption{Total R\&D compute and AI agent time horizon for OpenAI, 2018--2025.} 
  \label{fig:horizoncompute}
\end{figure}

Figure \ref{fig:horizoncompute} shows both trends for OpenAI. The time horizon data is taken directly from \cite{kwa2025measuring}. The compute estimate comes from first constructing a time series of OpenAI R\&D compute spend from their IRS Form 990s and \cite{EpochAIModels2025}. The compute estimates include training and experimental compute usage but not, for example inference compute.\footnote{\cite{EpochAIModels2025} already separates inference and R\&D compute, and for 2018 and 2022 we assume 0 inference costs. } We then use the methodology of \cite{cottier2024rising} to convert from compute spend in dollars to FLOPs.\footnote{More precisely, we use the inverse methodology, as they convert from FLOPs to dollars, while we convert from dollars to FLOPs. This conversion involves guesswork on which GPUs were used, the utilization rates and the prices, but it should give a roughly accurate picture. }

Both trends are approximately straight lines on a log plot, indicating exponential growth. \footnote{Interestingly, the lines have nearly the same slope, indicating that compute and time horizon are growing at the same rate. The theory will not require this.} Combined with these stylized facts, our theoretical model will imply that the growth rate of time horizon is causally proportional to the growth rate of compute. 

\subsection{Setup}\label{sec:setup}
Let 
\begin{itemize}
    \item $Y$ = METR's 50\% accuracy time horizon in minutes,
    \item $C$ = training compute spend in FLOPs
    \item $E$ = experimental compute spend in FLOPs,
    \item $A$ = algorithmic multiplier that scales training compute to effective compute,
    \item $t$ = calendar time in years. 
\end{itemize}

We start with two equations. First, we assume there is some function $F$ that maps effective training compute to time horizon. We make no assumptions on $F$ except differentiability:
\begin{equation}
    Y = F(AC).
    \label{eq:Y-FAC}
\end{equation}

Next, we assume algorithmic progress follows the standard economic model of ideas \citep{jones1995r}: 
\begin{equation}
    \frac{dA}{dt} \propto A^{1-\beta} E^\lambda,\qquad \lambda, \beta > 0.
    \label{eq:dA}
\end{equation}
This equation says that the rate of algorithmic progress is governed by two factors: $A^{1-\beta}$, which denotes that over time ideas may get harder to find, and $E$, which denotes how much compute is dedicated to researching algorithms. 

In this model, we assume that the effective inputs to algorithmic progress are well-approximated by experimental compute only. This could be justified if the primary bottleneck to progress is compute, not research labor. The primary reason we make this assumption is that it is plausible and it simplifies the relation between the compute growth and time horizon.\footnote{\Cref{sec:adding-labor} considers relaxing this assumption by adding a human researcher labor term.} Note that this explicitly assumes that a software-only singularity is not possible because labor (either human or AI provided) is not the constraint to more algorithmic progress. Therefore, our model is accurate only if the compute bottleneck approximation is good, or up to the point in time at which a software-only singularity kicks in. 

\subsection{Theoretical Results}\label{sec:theory-results}
Using \eqref{eq:Y-FAC} and \eqref{eq:dA}, we link the pace of time horizon growth to that of compute growth. Recall that the growth rate of a variable is equal to its log derivative with respect to time. So our first step is to log-differentiate time horizon: 
\begin{align}
\frac{d\log Y}{dt}
&= \frac{1}{Y}\,\frac{dY}{dt} \notag\\
&= \frac{1}{Y}\,F'(AC)\!\left( A\frac{dC}{dt} + \frac{dA}{dt}\,C \right) \notag\\
&= \frac{C}{Y}AF'(AC)\,\!\left( \frac{d\log C}{dt} + \frac{d\log A}{dt} \right) \notag\\
&= \frac{C}{Y}\frac{\partial Y}{\partial C}\bigg|_{A}\,\!\left( \frac{d\log C}{dt} + \frac{d\log A}{dt} \right) \notag\\
&= \frac{\partial \log Y}{\partial \log C}\bigg|_{A}
   \left( \frac{d\log C}{dt} + \frac{d\log A}{dt} \right).
\label{eq:last}
\end{align}
This equation says the growth rate of time horizon is equal to the growth rate of training compute plus the growth rate of algorithms, all multiplied by the log derivative of time horizon with respect to compute.

To simplify this expression further, we can note that equation 
\eqref{eq:dA} relates algorithm growth to compute investment. Furthermore, after enough time, the growth rate of equation \eqref{eq:dA} converges to 
\begin{equation}
    \frac{d \log A}{dt} = \frac{\lambda}{\beta} \frac{d \log E}{dt}.
    \label{eq:ss}
\end{equation}
While there are transition dynamics to this steady state, compute is currently growing extremely quickly, so the half-life of convergence is on the order of months.\footnote{When we consider compute scaling slowdowns, the half-life will be slightly longer, but still fairly short while the inputs grow so fast.} Therefore, it is without much loss to ignore transition dynamics and work with equation \ref{eq:ss} directly.\footnote{All further results are therefore implicitly asymptotic. For simplicity, the theory assumes instantaneous transitions. In reality there must be some delay between obtaining compute, algorithmic progress, and having deployed models, if only due to the time allocated to model training.}

Plugging equation \eqref{eq:ss} into equation \eqref{eq:last} yields 
\begin{equation}
    \frac{d\log Y}{dt}
    = \frac{\partial \log Y}{\partial \log C}\bigg|_{A}
   \left( \frac{d\log C}{dt} + \frac{\lambda}{\beta}\frac{d\log E}{dt} \right)
    \label{eq:main2}
\end{equation}
This equation relates the growth rate of time horizon with the growth-rate of training compute plus the growth rate of experimental compute weighted by a few other terms. 

Notice that $E + C$ denotes R\&D compute spend. If we are willing to assume that training compute and experimental compute are proportional to total R\&D spend\footnote{In Appendix \Cref{sec:training-and-experimental-compute} we show what changes when we drop this assumption.}, then we can simplify this equation to our main equation of this note:

\begin{equation}
    \frac{d\log Y}{dt}
    = \frac{\partial \log Y}{\partial \log C}\bigg|_{A}
   \left( 1 + \frac{\lambda}{\beta} \right) \frac{d \log (E+C)}{dt}
    \label{eq:main}.
\end{equation}

From this equation, we can derive a few interesting results. First, because the stylized facts imply that the growth rate of time horizon and compute are constant, we can derive that $\dfrac{\partial\log Y}{\partial\log C}\bigg|_{A}$ is constant, i.e., it doesn't itself depend on algorithms or compute. 
\begin{lemma}\label{lem:const}
    \Cref{fact:time,fact:compute} $\implies \dfrac{\partial\log Y}{\partial\log C}\bigg|_{A}$ is constant.
\end{lemma}
\noindent \textit{Proof.} See \Cref{proof:lemma-const}.

Second, this implies that the growth rate of time horizon is proportional to the growth rate of total R\&D compute. 
\begin{prop}\label{prop:linear}
    \Cref{fact:time,fact:compute} $\implies$ $\dfrac{d\log Y}{dt} = c \,\dfrac{d\log (E+C)}{dt}$, where $c \in \mathbb{R}$ and $\newline c = \dfrac{\partial\log Y}{\partial\log C}\bigg|_{A}(1 + \frac{\lambda}{\beta})$
\end{prop}
\noindent \textit{Proof.} See \Cref{proof:prop-linear}.

Because $c = \dfrac{\partial\log Y}{\partial\log C}\bigg|_{A}(1 + \frac{\lambda}{\beta})$, the substantive implication is that $\dfrac{\partial\log Y}{\partial\log C}\bigg|_{A}$ is a constant. Notice $\dfrac{\partial\log Y}{\partial\log C}\bigg|_{A}$ is a property of $F$, the function that maps effective training compute to time horizon, as $\dfrac{\partial\log Y}{\partial\log C}\bigg|_{A}$ measures the returns for time horizon when increasing training compute. This term is constant if and only if $F$ is a power-law function. 

Our empirical results only speak to our data from 2019--2025, like any empirical study we cannot guarantee the relationship holds out of distribution after 2025. However, $F$ has exhibited power-law-like behavior over many orders of magnitude, so we believe it is reasonable to conclude that $F$ is a power-law function. If $F$ is indeed a power-law, we have a causal model that says the growth rate of time horizon and compute are proportional. Therefore, under this model, if the growth rate of total R\&D compute drops by half, then the growth rate of time horizon will also drop by half.

\section{Empirics}\label{sec:empirics}

In this section, we test an empirical prediction of the theory as a sanity check on our results. In particular, we study whether \Cref{lem:const} holds, i.e., that $\dfrac{\partial\log Y}{\partial\log C}\bigg|_{A}$ is constant\footnote{Note that the observations that follow will not provide distinguishing evidence for our model relative to other possible models for which $\dfrac{\partial\log Y}{\partial\log C}\bigg|_{A}$ may be constant.}. The following proposition gives us a simple test for this. 

\begin{prop}\label{prop:iff}
    $\dfrac{\partial\log Y}{\partial\log C}\bigg|_{A}$ is constant $\iff \log Y=\alpha+\gamma\log A+\gamma\log C$.
\end{prop}
\noindent \textit{Proof.} See \Cref{proof:prop-iff}.

The equation $\log Y=\alpha+\gamma\log A+\gamma\log C$ implies that, for any fixed algorithm, $\log Y$ vs.\ $\log C$ should be linear. Therefore, we test this by taking families of language models that have the same algorithms but with varying compute. We selected the Llama~3.1 and Qwen~2.5 families, since these models have multiple versions with varying compute, while the training algorithms are held constant (there are minimal distillation or data-quality changes). We measure $Y$ via imputation from benchmarks using the methodology from \citet{metr_time_horizon_domains_2025}, except for the Llama family, where we recorded the METR time horizon on the METR task suite directly.\footnote{We exclude Qwen~2.5 0.5B for GPQA, as 1.5B approaches the guessing bound on the benchmark.} However, this direct measurement 
was conducted somewhat less carefully than is standard for METR’s time horizon evaluations and should therefore be given lower weight.

\begin{figure}[H]
  \centering
  \includegraphics[width=\textwidth]{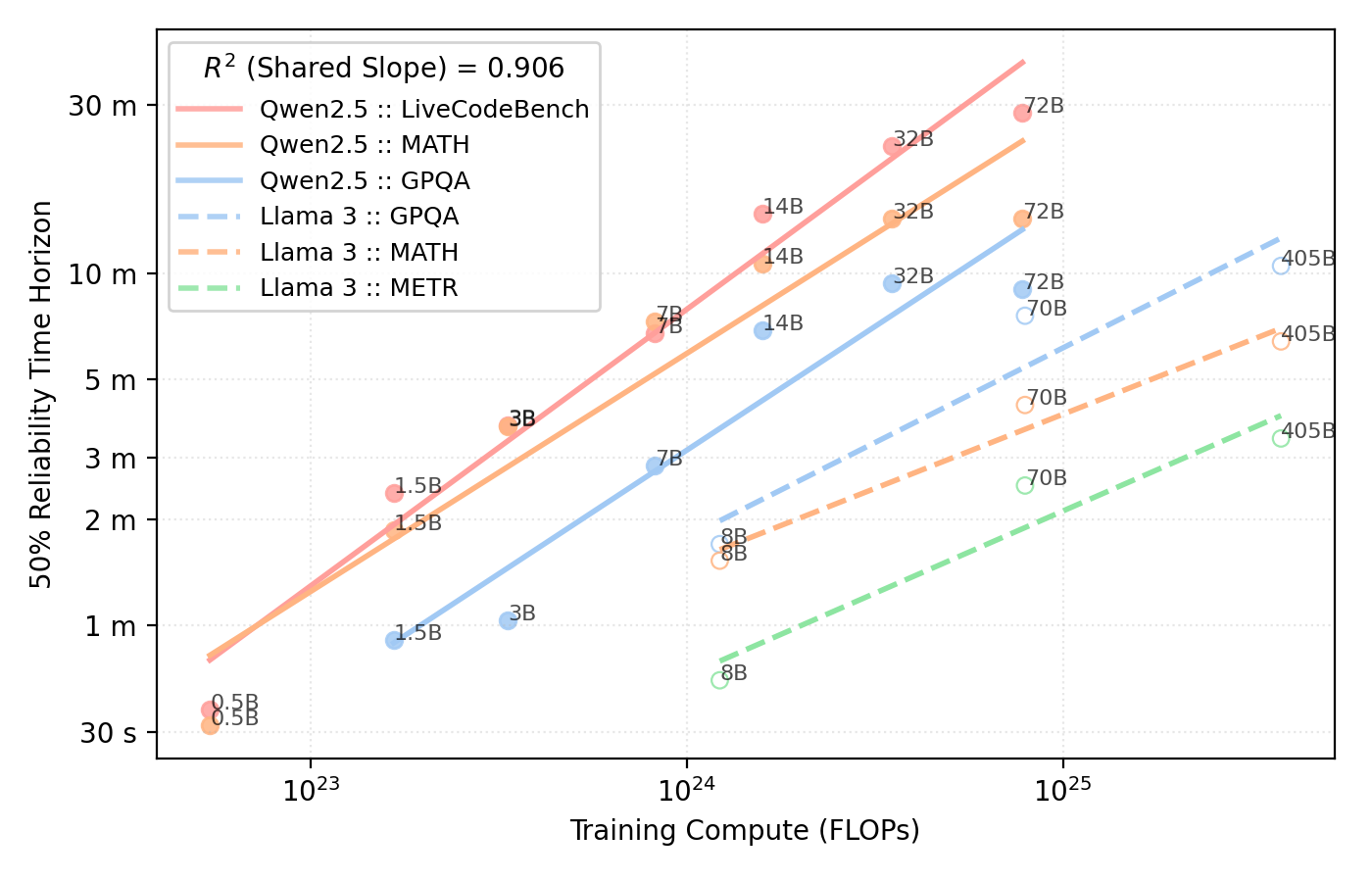}
  \caption{Log-log fit of time horizon vs.\ compute.}
  \label{fig:standard}
\end{figure}

\Cref{fig:standard} shows the fit. The slopes across models and benchmarks are fairly similar, and the linear fit appears reasonable, albeit with a leveling off at the top end. To summarize the results quantitatively, we estimate the equation 
\[
    \text{log time horizon} = \text{constants for each model-benchmark} + \beta \text{log training compute}
\]
and report the partial $R^2$ that the model achieves versus just the constants. We call this the shared slope $R^2$ as we are only fitting a single slope for all $30$ data points. The shared slope $R^2$ would be 1 if and only if all model-benchmarks have the same log time horizon vs. log training compute slope and a linear fit was perfect. We obtain a shared slope $R^2$ of $0.906$. An equivalent and visual way to evaluate this shared slope is to residualize both time horizon and training compute by model-benchmark and compute the $R^2$ in the standard way, which is done in Appendix \Cref{sec:sharedslopesapp}. 

One possible explanation for the leveling off at the top in \Cref{fig:standard} is that compute-optimality is not held fixed. Llama~3, for example, is trained on the same pre-training corpus regardless of model size: 8B $\to$ 70B $\to$ 405B just increases the corresponding number of parameters. This is plausibly different from what has happened for models in the METR sample, where the ratio of training data to parameters has likely been held closer to constant. To adjust for this effect, we replot the same relationship, but instead of using raw compute, we use the amount of compute needed to reach the same loss if the training were compute-optimal.\footnote{The important part is not that the models in the METR sample were themselves compute-optimal, but that the degree of compute optimality was being held constant. } We make this adjustment using the Chinchilla loss formula \citep{hoffmann2022trainingcomputeoptimallargelanguage} to calculate the training loss for each model given its training dataset size and number of parameters. We then calculate the minimum amount of compute needed to reach that same loss. 

\begin{figure}[H]
  \centering
  \includegraphics[width=\textwidth]{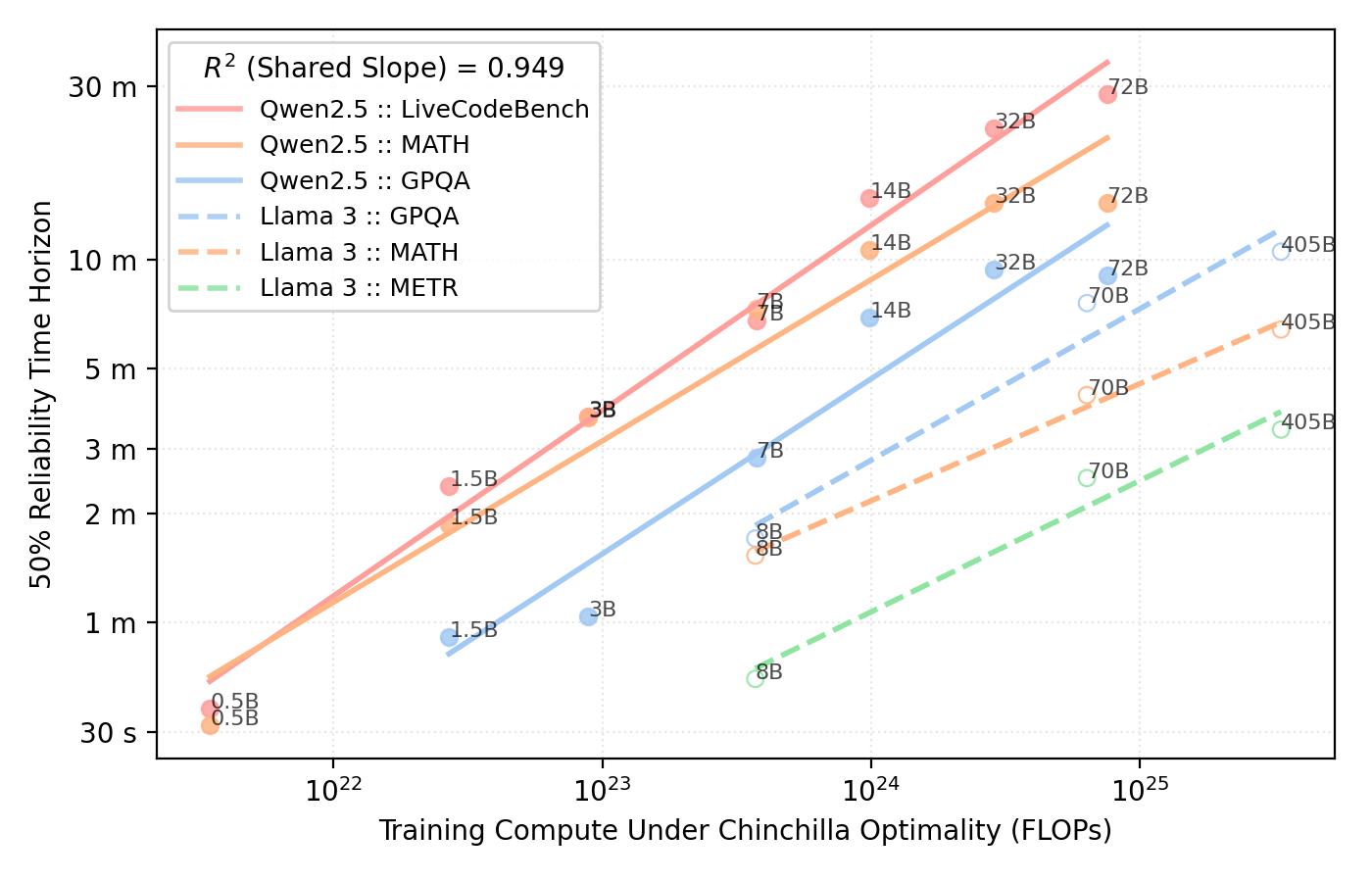}
  \caption{Re-estimated log-log fit, using compute adjusted for compute-optimality.}
  \label{fig:compute-opt}
\end{figure}

The resulting fit (\Cref{fig:compute-opt}) is better, with the shared slope $R^2$ increasing to $0.949$. 

Overall, the linear fit (in log space) is not perfect, but is reasonably accurate over about 4 orders of magnitude of training compute. Therefore, our interpretation is that the experimental evidence is broadly consistent with the theory, but since it comes from only 2 model scale-ups and there is some evidence of concavity, we do not consider it decisive evidence. However, for our forecasts to be accurate, we need $\dfrac{\partial\log Y}{\partial\log C}\bigg|_{A}$ estimated in 2018--2025 to be a good estimate of $\dfrac{\partial\log Y}{\partial\log C}\bigg|_{A}$ from 2025 onwards. If the linear fit was perfect, this would imply that these terms are exactly equal, but even with some error our forecasts would still be reasonably accurate. Appendix Section \ref{sec:concavity} checks the robustness of our forecasts by allowing our estimate of $\dfrac{\partial\log Y}{\partial\log C}\bigg|_{A}$ to fall as $C$ increases. 

\section{Forecasting}\label{sec:forecasting}
Proposition \ref{prop:linear} gives us the following relationship between compute growth and time horizon growth.
\begin{equation}
    \text{future horizon growth}
    = \frac{\text{past horizon growth}}{\text{past compute growth}}
      \cdot \text{future compute growth}
    \label{eq:horizon-growth}
\end{equation}
We can estimate the past growth rate ratio from the data in Figure \ref{fig:horizoncompute}. We subset this data to only OpenAI, as they have the largest data coverage on both time horizon and compute. Once we have this estimated, equation \ref{eq:horizon-growth} lets us convert from an implied growth rate change in compute to an implied growth rate change in time horizon. 

To get an estimate of implied future compute growth, we use OpenAI's own forecasts of their compute spend up to 2030 as reported in \cite{EfratiMuppidi2025OpenAI350B} and then assume compute growth stabilizes at the 2029-2030 rate. To account for computing improvements, we assume that FLOPs per \$ improves at the same rate per year it did from 2018--2025. 

\begin{figure}[H]
  \centering
  \includegraphics[width=\textwidth]{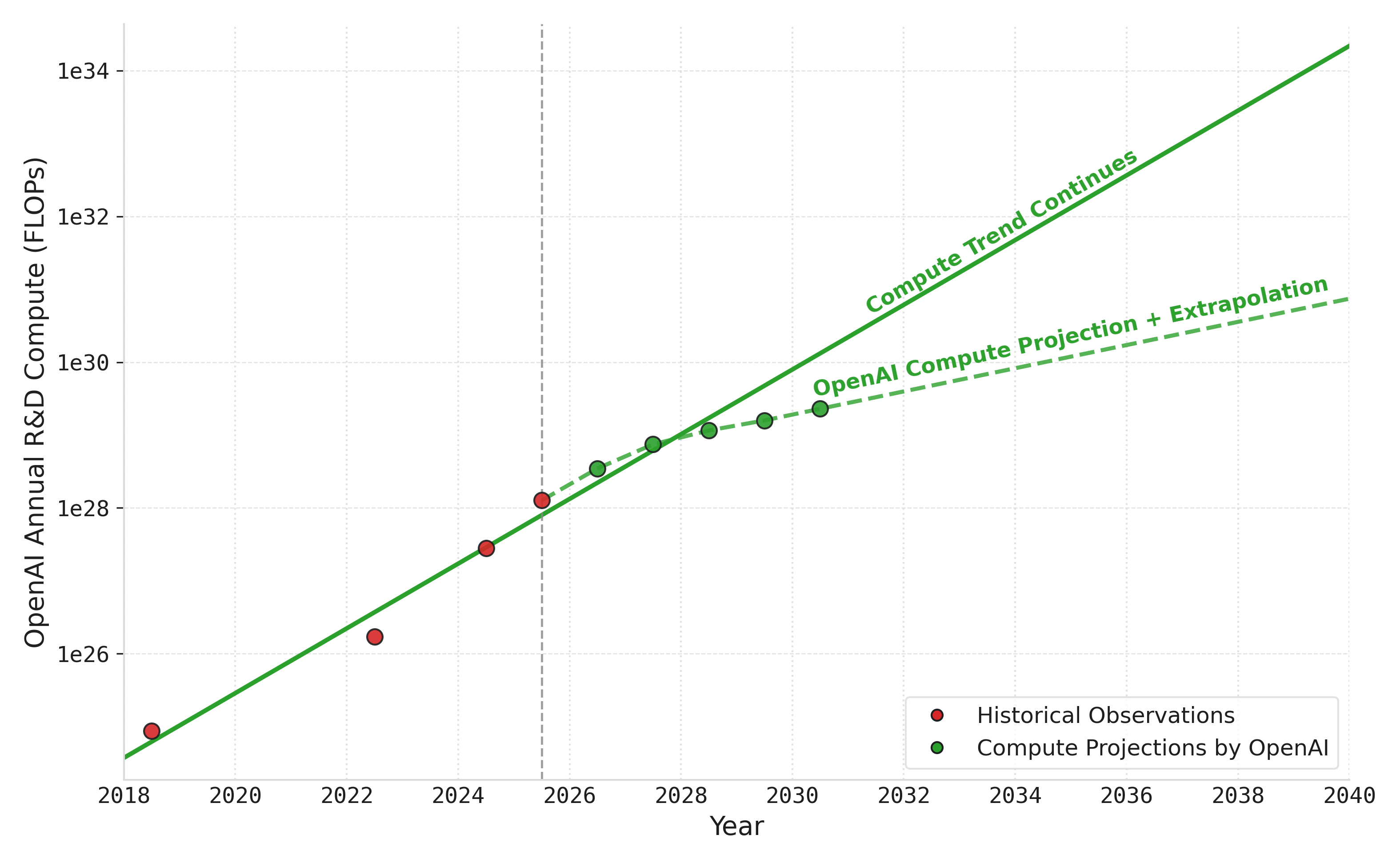}
  \caption{OpenAI projected compute.}
  \label{fig:openaicomputepath}
\end{figure}

Figure \ref{fig:openaicomputepath} plots the projected compute path, along with the historical path we described in previous sections. The `Compute Trend Continues' line shows estimated trend of compute if OpenAI compute continues to grow at the 2018--2025 rate, while the `OpenAI Compute Projection + Extrapolation'  grows at a similar rate until about 2028 at which point compute growth slows.\footnote{Note the difference in levels from 2025-2027 is due to `Compute Trend Continues' starting from the trend line since it is a linear extrapolation, while `OpenAI Compute Projection + Extrapolation' starts from the last point since it is a piecewise linear fit between points. This is not material for our time horizon forecast, which depends only on the slope. }

\begin{figure}[H]
  \centering
  \includegraphics[width=\textwidth]{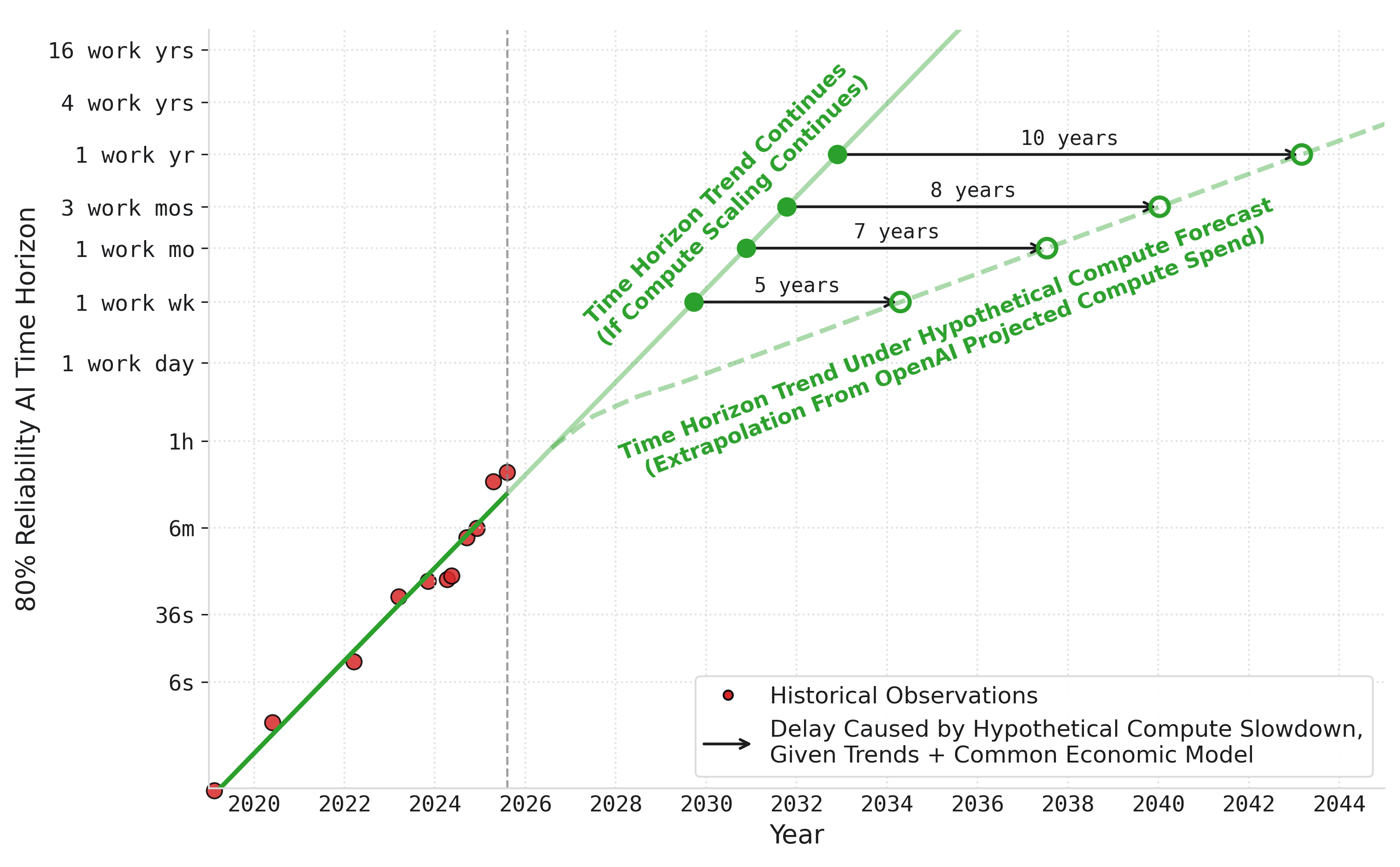}
  \caption{80\% reliability time horizon forecast.}
  \label{fig:p80}
\end{figure}

Figure \ref{fig:fig1} shows the corresponding $50\%$ time horizon path. Figure \ref{fig:p80} shows the corresponding $80\%$ time horizon path. 

\begin{figure}[H]
  \centering
  \includegraphics[width=\textwidth]{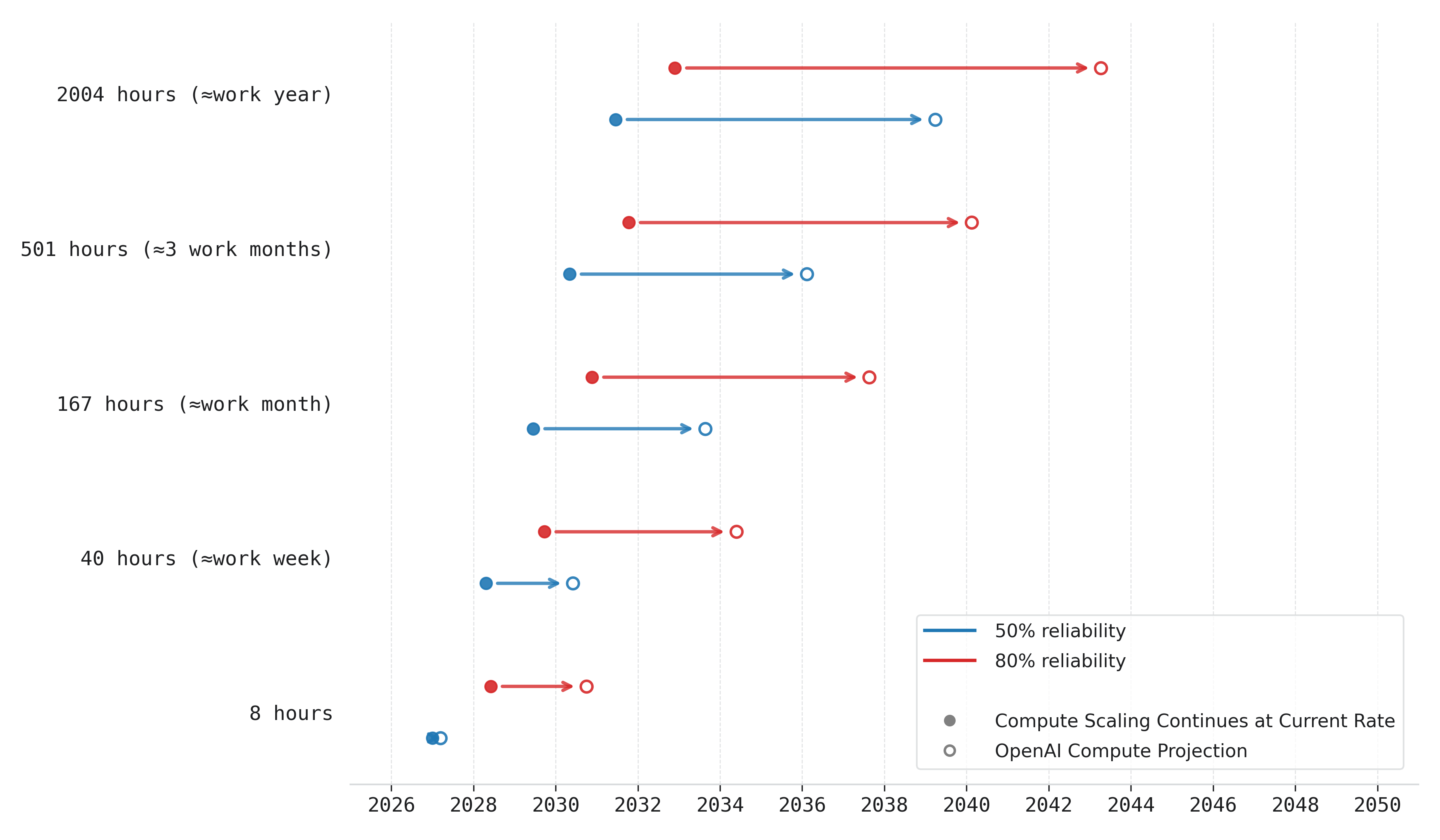}
  \caption{Implied milestone delay from compute slowdown}
  \label{fig:delay}
\end{figure}

Notice that Figure \ref{fig:p80} has longer delays than Figure \ref{fig:fig1}. The intuition for this is that because $50\%$ time horizon is ahead of $80\%$ time horizon, the $50\%$ time horizon gets to spend more relative time in the current fast growth mode than the slow growth mode. This explanation is generally why we see longer delays for higher time horizons and higher reliability. Figure \ref{fig:delay} captures this dynamic in a graph, where we display the delay introduced for various time horizons at various reliabilities. 

\subsection{Webapp}
We release a web app \url{https://timehorizonforecast.com/} where users can convert any compute scaling path to the corresponding path for METR time horizon. The compute path can either be supplied in FLOPs directly or in 2025 USD.\footnote{If the path is supplied in USD, we use the same technique described previously to convert to FLOPs, where we assume that FLOPs per dollar improves at the 2018--2025 rate.} 
\begin{figure}[H]
    \centering
    \includegraphics[width=\textwidth]{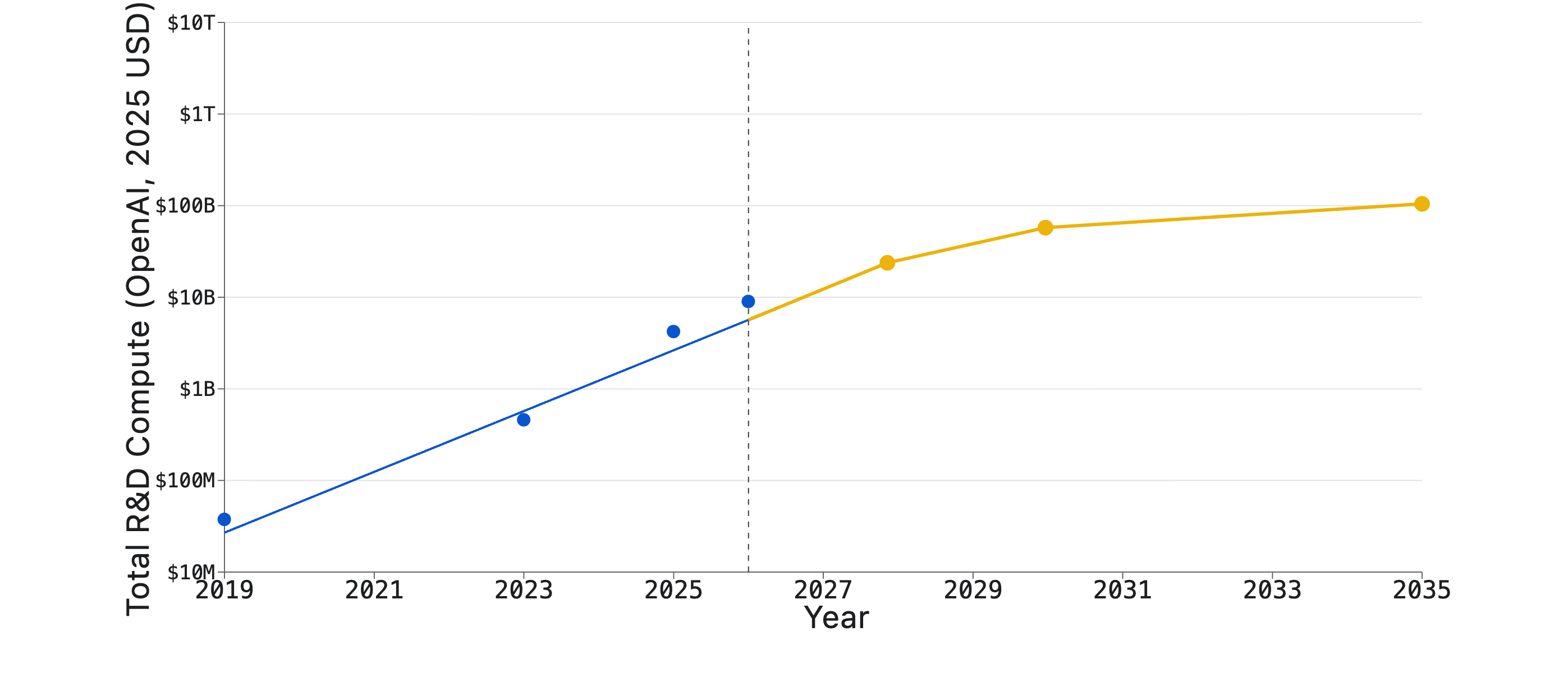}
    \caption{Example compute path in USD on the Web App.}
    \label{fig:computepath}
\end{figure}

\begin{figure}[H]
    \centering
    \includegraphics[width=\textwidth]{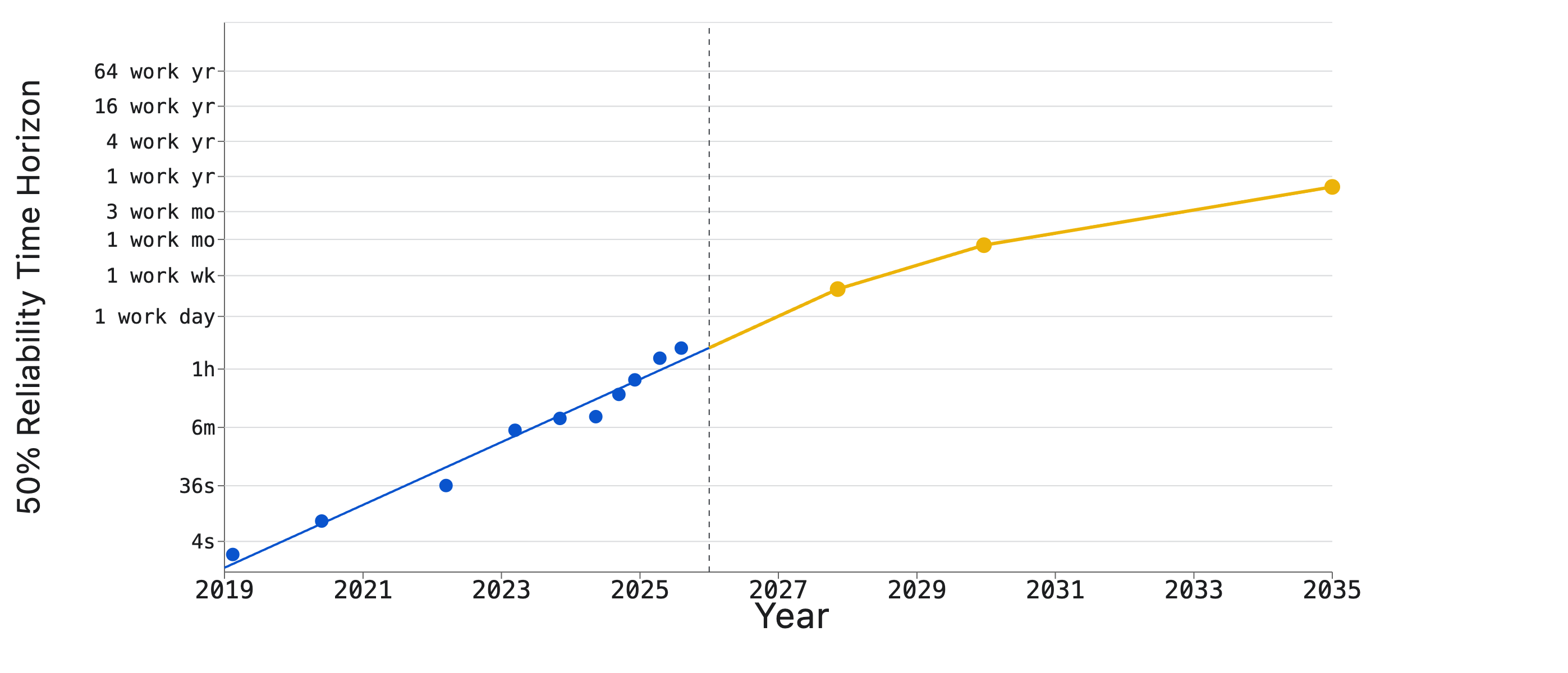}
    \caption{Example METR time horizon path implied by compute path in \Cref{fig:computepath}.}
    \label{fig:timepath}
\end{figure}

\newpage
\clearpage
\appendix

\section{Author contributions}

\textbf{Parker Whitfill} produced theoretical results and contributed the large majority of data analysis and draft writing. \textbf{Ben Snodin} collected time horizon data necessary for the figures in section \ref{sec:empirics}. \textbf{Joel Becker} gave guidance on the project and contributed to data analysis and draft writing.

\section{Proofs}\label{app:proofs}

\subsection{Proof of Lemma~\ref{lem:const}}\label{proof:lemma-const}
Rearranging \eqref{eq:main} yields
\begin{equation}
\frac{\partial \log Y}{\partial \log C}\bigg|_{A}
= \frac{\dfrac{d\log Y}{dt}}{\dfrac{d\log C}{dt}\,\Bigl(1+\dfrac{\lambda}{\beta}\Bigr)}.
\label{eq:rearranged}
\end{equation}
\Cref{fact:time} implies $\dfrac{d\log Y}{dt}$ is constant and \Cref{fact:compute} combined with our constant training compute proportion assumption implies $\dfrac{d\log C}{dt}$ is constant. Therefore, the entire right-hand side of \eqref{eq:rearranged} is constant. Hence the left-hand side is constant. \qed

\subsection{Proof of Proposition~\ref{prop:linear}}\label{proof:prop-linear}
By \Cref{lem:const}, $\dfrac{\partial\log Y}{\partial\log C}\bigg|_{A}$ is constant. Therefore, let 
\[
c = \frac{\partial \log Y}{\partial \log C}\bigg|_{A}\!\left(1 + \frac{\lambda}{\beta}\right).
\]
Using \eqref{eq:main}, we have 
\[
\frac{d\log Y}{dt} = c\,\frac{d\log (E+C)}{dt}.
\]
\qed

\subsection{Proof of Proposition~\ref{prop:iff}}\label{proof:prop-iff}
($\Rightarrow$)
Recall, 
\[
\frac{\partial \log Y}{\partial \log C}\bigg|_{A}
=\frac{AC\,F'(AC)}{F(AC)}
=\frac{d\log F(AC)}{d\log(AC)}.
\]
Therefore, if $\dfrac{\partial\log Y}{\partial\log C}\bigg|_{A}\equiv\gamma$ is constant, then
\[
\frac{d\log F(AC)}{d\log(AC)}=\gamma.
\]
Integrating with respect to \(\log(AC)\) gives
\[
\log F(AC)=\alpha+\gamma\log(AC),
\]
hence \(\log Y=\alpha+\gamma(\log A+\log C)\).

($\Leftarrow$) If \(\log Y=\alpha+\gamma\log A+\gamma\log C\), then holding \(A\) fixed yields
\[
\frac{\partial \log Y}{\partial \log C}\bigg|_{A}=\gamma.
\]
\qed

\section{Adding Labor to Algorithmic Progress}\label{sec:adding-labor}
In the main text, we assumed that the only input to algorithmic progress research is experimental compute $E$, as in equation~\eqref{eq:dA}. Here we briefly sketch how the results might change if algorithmic ideas also depend on human researcher labor $L$.

Suppose instead that the Jones law of motion has as input a Cobb-Douglas production function between compute and labor. 
\begin{equation}
    \frac{dA}{dt}
    \propto A^{1-\beta}\bigl(E^{1-\alpha}L^{\alpha}\bigr)^{\lambda},
    \qquad 0 < \alpha < 1,\ \lambda,\beta > 0.
    \label{eq:dA-EL}
\end{equation}
As before, after enough time the log derivative of $A$ with respect to $t$ converges to
\begin{equation}
    \frac{d\log A}{dt}
    = \frac{\lambda}{\beta}
      \left(
        (1-\alpha)\frac{d\log E}{dt}
        + \alpha\frac{d\log L}{dt}
      \right).
    \label{eq:ss-EL}
\end{equation}

Substituting equation \eqref{eq:ss-EL} into equation~\eqref{eq:last} gives
\begin{equation}
    \frac{d\log Y}{dt}
    = \frac{\partial \log Y}{\partial \log C}\bigg|_{A}
      \left[
        \frac{d\log C}{dt}
        + \frac{\lambda}{\beta}
          \left(
            (1-\alpha)\frac{d\log E}{dt}
            + \alpha\frac{d\log L}{dt}
          \right)
      \right].
    \label{eq:Y-with-E}
\end{equation}

Maintaining the assumption from the main text that experimental compute and training compute grow at similar rates, then we get
\begin{align}
    \frac{d\log Y}{dt}
    &\approx \frac{\partial \log Y}{\partial \log C}\bigg|_{A}
      \left[
        \frac{d\log (E + C)}{dt}
        + \frac{\lambda}{\beta}
          \left(
            (1-\alpha)\frac{d\log (E + C)}{dt}
            + \alpha\frac{d\log L}{dt}
          \right)
      \right] \notag\\[4pt]
    &= \frac{\partial \log Y}{\partial \log C}\bigg|_{A}
      \left[
        \Bigl(1 + \frac{\lambda}{\beta}(1-\alpha)\Bigr)\frac{d\log (E + C)}{dt}
        + \frac{\lambda}{\beta}\alpha\,\frac{d\log L}{dt}
      \right].
    \label{eq:Y-CL-final}
\end{align}

First, notice that if $\frac{d\log (E + C)}{dt}$ and $\frac{d\log L}{dt}$ grow at similar rates, then we get the same equation from the main text. 
This just says that if compute slowdowns are paired with labor slowdowns, then the main text equation is accurate. This is plausible because both are driven by the same underlying investment process. 

Second, even if we model compute slowdowns as separate from labor slowdowns, we get fairly similar results. \cite{epoch2025thesoftwareintelligenceexplosiondebateneedsexperiments} estimate that at OpenAI, $\alpha \approx .67$, $\frac{\lambda}{\beta} \approx .95$, and we estimate $\frac{\partial \log Y}{\partial \log C}\bigg|_{A} \approx .454$ in the main paper. Finally, \citep{EpochAIModels2025} finds that OpenAI staff has been $1.6x$-ing a year. Putting these terms together, a back of the envelope calculation implies that dropping compute growth by half would drop time horizon growth by $\approx 45\%$ instead of the full $50\%$. Therefore, while including staff does break the perfect proportionality result, the predictions do not change by much. 

\section{Training And Experimental Compute}\label{sec:training-and-experimental-compute}

We obtained equation \eqref{eq:main} from equation \eqref{eq:main2} by assuming that $C$ and $E$ move proportionally with total R\&D compute $E+C$. Here we show exactly what changes when we drop that assumption.

Write the shares
\[
s_C:=\frac{C}{E+C},
\qquad
s_E:=\frac{E}{E+C}.
\]
Then the following identity holds:
\begin{equation*}\label{eq:error-decomp-logs}
\left(\frac{d\log C}{dt}+\frac{\lambda}{\beta}\frac{d\log E}{dt}\right)
=
\left(1+\frac{\lambda}{\beta}\right)\frac{d\log (E+C)}{dt}
\;+\;
\underbrace{\Bigl(s_E-\tfrac{\lambda}{\beta}s_C\Bigr)
\left(\frac{d\log C}{dt}-\frac{d\log E}{dt}\right)}_{:=\ \varepsilon_t}.
\end{equation*}
Plugging into equation \eqref{eq:main2} gives
\begin{equation}\label{eq:main2-exact-logs}
\frac{d\log Y}{dt}
=
\frac{\partial \log Y}{\partial \log C}\bigg|_{A}
\left[
\left(1+\frac{\lambda}{\beta}\right)\frac{d\log (E+C)}{dt}
\;+\;
\varepsilon_t
\right].
\end{equation}

When is the extra term small or zero? 
\begin{itemize}
\item If $d\log C/dt \approx d\log E/dt$. This is the constant proportions of total R\&D compute assumption we make in the main paper. 
\item If $s_E \approx \frac{\lambda}{\beta} s_C$. Since the returns to growing $E$ are $\frac{\lambda}{\beta}$, intuitively this says $s_E$ divided by its return to effective compute equals $s_C$ divided by its return to effective compute $(1)$. This would occur in a simple Cobb-Douglas model, although it does not formally occur here without further assumptions. 
\end{itemize}
Both conditions seem plausible and if either of them is true then we have zero error. Moreover, the error is proportional to the differences of these terms, so even if equality does not hold exactly in practice, it may hold approximately. 

Since total R\&D compute is necessarily a sum of experimental and training compute, we suspect that using this as the input measure is relatively accurate. We suspect that this is more accurate than using training compute estimates from \citep{epoch2024trainingcomputeoffrontieraimodelsgrowsby45xperyear} and then inferring experimental compute by the proportionality assumption. 

\section{Shared Slopes}\label{sec:sharedslopesapp}
In \Cref{sec:empirics} we reported the shared slope $R^2$. Here we show visually what this fit corresponds to. We residualize both log time horizon and log training compute by model-benchmark. For example, residualized log time horizon equals log time horizon minus the mean log time horizon among the model-benchmark group. By the Frisch-Waugh theorem, this is equivalent to regressing log time horizon on a single log training compute slope term and model-benchmark constants. 

\begin{figure}[H]
  \centering
  \includegraphics[width=\textwidth]{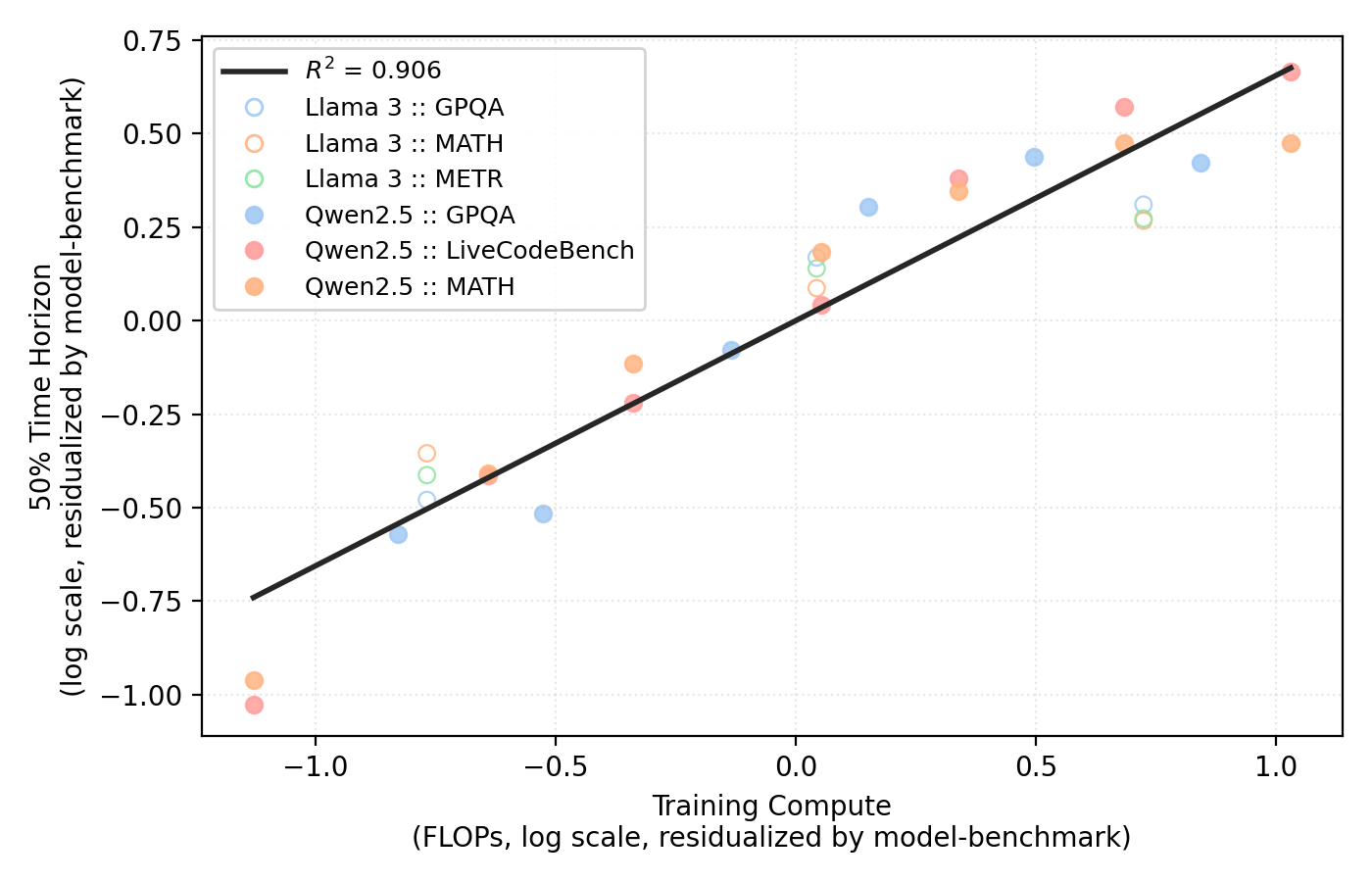}
  \caption{Residualized version of \cref{fig:standard}.}
  \label{fig:resid1}
\end{figure}

\begin{figure}[H]
  \centering
  \includegraphics[width=\textwidth]{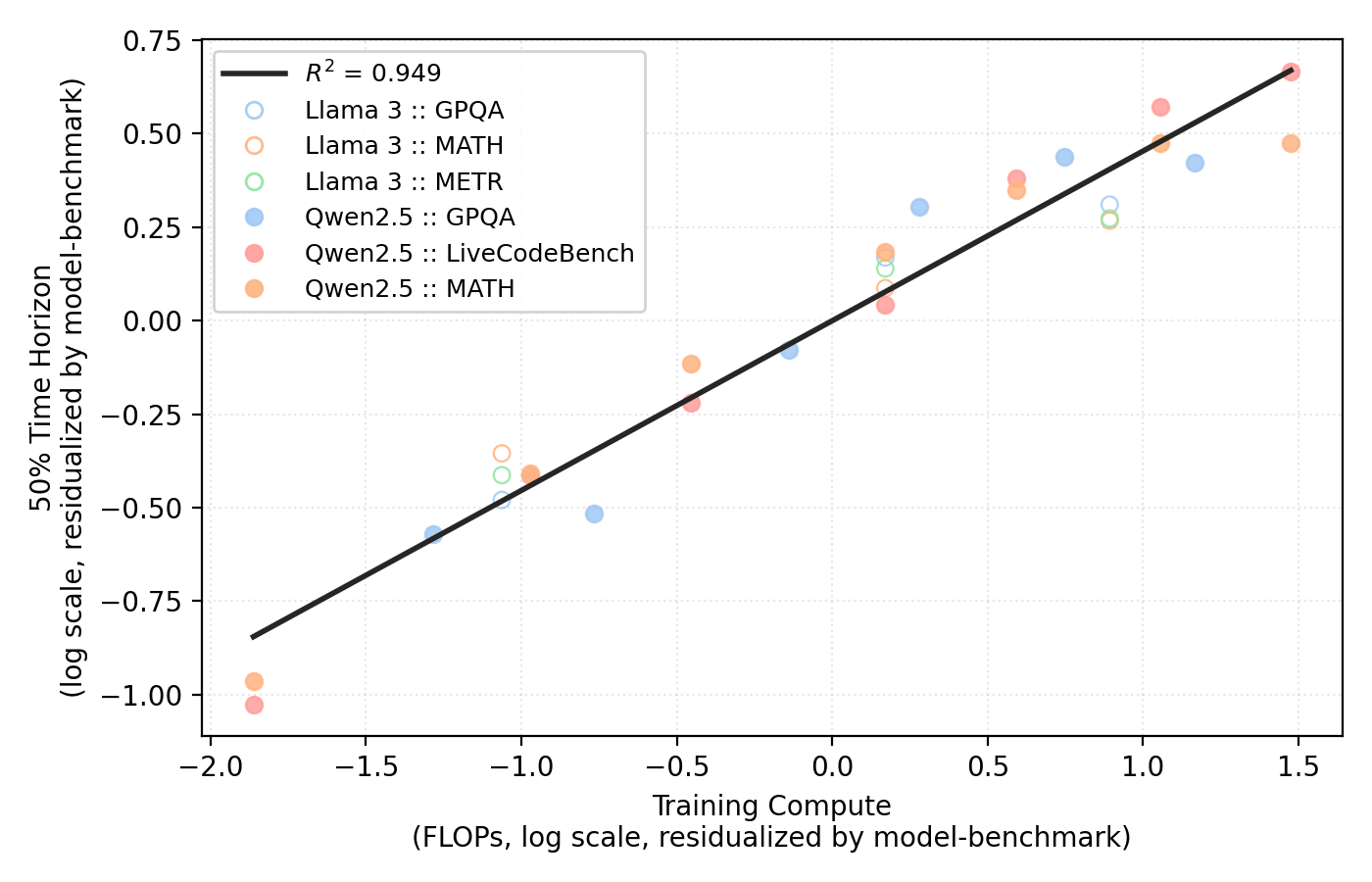}
  \caption{Residualized version of \Cref{fig:compute-opt}.}
  \label{fig:resid2}
\end{figure}

\section{Forecasting Time Horizon Under Concavity}\label{sec:concavity}
Our results in \Cref{fig:compute-opt} could reasonably be viewed as weak evidence that $\frac{\partial \log Y}{\partial \log C}\bigg|_{A}$ declines over time. As a robustness check to our main results, we go back to the same data in Figure \ref{fig:compute-opt} and estimate via NLS: 
\begin{align*}
    \log Y = \text{model-benchmark constants} + 
    \beta(\frac{X^\rho -1}{\rho})
\end{align*}
In this model, $\frac{\partial \log Y}{\partial \log C}\bigg|_{A} = \beta C^\rho$, which generalizes from the $\rho = 0$ case in the main text. This more flexible form allows us to fit slightly better, with a partial $R^2$ of $0.972$. Moreover, our estimate $\rho < 0$ so $\frac{\partial \log Y}{\partial \log C}\bigg|_{A}$ declines over time, giving us concavity. 

To account for this concavity in our forecast, we return to equation \ref{eq:main} and substitute in our estimated elasticity 
\begin{align*}
    \frac{d\log Y}{dt}
    & = \frac{\partial \log Y}{\partial \log C}\bigg|_{A}
   \left( 1 + \frac{\lambda}{\beta} \right) \frac{d \log (E+C)}{dt} \\
   & = \beta C^\rho \left( 1 + \frac{\lambda}{\beta} \right) \frac{d \log (E+C)}{dt}
\end{align*}
Given our path of total R\&D compute from 2018--2025, we then estimate $\frac{\lambda}{\beta}$ to make the predicted values for $\log Y$ most closely match what we observe in the data.\footnote{Note that, because of the concavity, we explicitly assume how training compute and total compute relate: we assume training compute is $\frac{1}{10}$ following \cite{epoch2025openaicomputespend}.}

\begin{figure}[H]
  \centering
  \includegraphics[width=\textwidth]{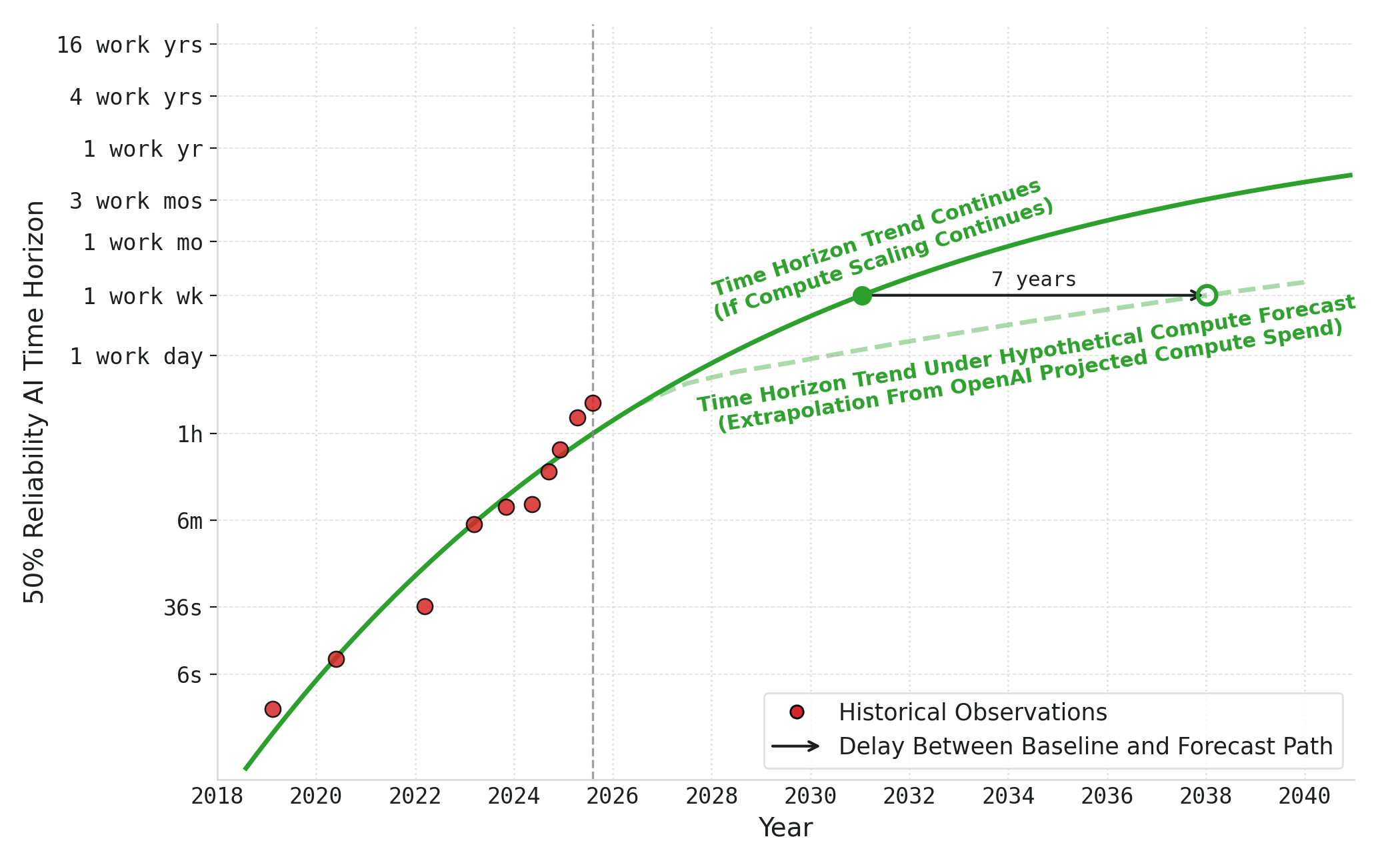}
  \caption{Concave Version of Figure \ref{fig:fig1}}
  \label{fig:concave}
\end{figure}

Figure \ref{fig:concave} shows a version of Figure \ref{fig:fig1} that incorporates concavity in $F$. Interestingly, time-horizon milestone delays lengthen, with 1-week time horizon being delayed by 7 years instead of 2 years in Figure \ref{fig:fig1}. The intuition is that introducing concavity means the compute needed to reach a given capability level rises as you need to overcome the decreasing returns. We discussed earlier in the paper that harder capability levels are more delayed by a compute slowdown. Moreover, this effect bites harder the more concave the returns are. Therefore, our main results under no concavity provide a lower bound on the delay introduced if there is some concavity.\footnote{Symmetrically, if returns are increasing over time, then the main results are an upper bound. } 

\section{Algorithmic Progress}\label{sec:algorithmic-progress}
The framework also implies a back-of-the-envelope estimate of algorithmic progress. Starting from \eqref{eq:last},
\[
\frac{d\log Y}{dt}
= \frac{\partial \log Y}{\partial \log C}\bigg|_{A}
  \left( \frac{d\log C}{dt} + \frac{d\log A}{dt} \right),
\]
solve for the algorithmic progress rate:
\begin{equation}
\frac{d\log A}{dt}
= \frac{1}{\dfrac{\partial \log Y}{\partial \log C}\bigg|_{A}}
  \frac{d\log Y}{dt}
  - \frac{d\log C}{dt}.
\label{eq:gA-rearranged}
\end{equation}

We already estimated $\dfrac{\partial \log Y}{\partial \log C}\bigg|_{A}$ in Figure \ref{fig:compute-opt}, $\frac{d\log Y}{dt}$ can be directly estimated in the METR data, while $\frac{d\log C}{dt}$ can be estimated in the METR data by merging it with training compute estimates from \cite{EpochAIModels2025}. For these estimates, we consider all non-Chinese models in the METR data that have compute estimates, not just those from OpenAI. 

The following table shows the estimate of algorithmic progress we get from our data. 
\begin{table}[H]
  \centering
  \begin{threeparttable}
    \caption{Algorithmic progress implied by the METR p50\% horizon.}
    \label{tab:algo-progress-p50}
    \begin{tabularx}{\textwidth}{@{}>{\centering\arraybackslash}X>{\centering\arraybackslash}X@{}}
      \toprule
      $\mathrm{d}\ln A/\mathrm{d}t$ & 95\% CI \\
      \midrule
      1.265 & [0.667,\, 3.614] \\
      \bottomrule
    \end{tabularx}
    \begin{tablenotes}\footnotesize
      \item Confidence interval from 10000 paired METR resamples with Normal draws around the clustered elasticity estimate.
    \end{tablenotes}
  \end{threeparttable}
\end{table}

The point estimate corresponds to roughly $3.5\times$ per year, which closely matches the estimate in \cite{ho2024algorithmic}. One advantage of this estimate over \cite{ho2024algorithmic} is that it includes post-training and measures progress towards an economically relevant term, time horizon. However, because we estimate \(\dfrac{\partial \log Y}{\partial \log C}\bigg|_{A}\) using only Qwen-2.5 and Llama-3.1, we should be cautious about the external validity of these estimates.

\section{Acknowledgments}
Thanks to Hjalmar Wijk, Phil Trammell, Thomas Kwa, David Rein, Charles Foster, Nikola Jurkovic, Ryan Greenblatt, Anson Ho, Jacob Hilton, David Farhi, Tom Cunningham, Beth Barnes and others for helpful comments.


\end{document}